\documentclass[prb,twocolumn,showkeys,showpacs,amsmath,amsfonts,floatfix,superscriptaddress,nofootinbib]{revtex4-1} 
\usepackage{amssymb,amsmath,amstext}                
\usepackage{graphicx}                                               
\usepackage{tikz-network}
\usepackage{epstopdf}                                               
\usepackage{color}       
\usepackage{bm}
\usepackage{appendix}                                              
\usepackage[latin2]{inputenc}
\usepackage{ulem}
\usepackage{latexsym}
\usepackage{tabularx}
\usepackage{afterpage}
\usepackage{nonfloat}
\usepackage[colorlinks=true,citecolor=blue,linkcolor=magenta]{hyperref}
\usepackage{lipsum}

\def\be{\begin{equation}}
\def\ee{\end{equation}}
\def\bea{\begin{eqnarray}}
\def\eea{\end{eqnarray}}
\def\bi{\begin{itemize}}
\def\ei{\end{itemize}}

\newcommand{\braket}[2]{\mbox{$\langle #1  | #2 \rangle$}}

\newcommand{\changed}[1]{{\color{blue}{{#1}}}}

\begin{document}

\title{ 
Bang-bang preparation of quantum many-body ground states in two dimensions: \\
optimization of the algorithm with a two-dimensional tensor network
}

\author{Yintai Zhang}
\affiliation{Jagiellonian University, Doctoral School of Exact and Natural Sciences,
ul. \L{}ojasiewicza 11, 30-348 Krak\'ow, Poland
}
\affiliation{Jagiellonian University, 
             Faculty of Physics, Astronomy and Applied Computer Science,
             Institute of Theoretical Physics, 
             ul. \L{}ojasiewicza 11, 30-348 Krak\'ow, Poland }  
 
\author{Jacek Dziarmaga} 
\affiliation{Jagiellonian University, 
             Faculty of Physics, Astronomy and Applied Computer Science,
             Institute of Theoretical Physics, 
             ul. \L{}ojasiewicza 11, 30-348 Krak\'ow, Poland }  
\affiliation{Jagiellonian University, 
             Mark Kac Center for Complex Systems Research,
             ul. \L{}ojasiewicza 11, 30-348 Krak\'ow, Poland }

\date{January 4, 2024}

\begin{abstract}
A bang-bang (BB) algorithm prepares the ground state of a two-dimensional (2D) quantum many-body Hamiltonian $H=H_1+H_2$ by evolving an initial product state alternating between $H_1$ and $H_2$. We use the neighborhood tensor update to simulate the BB evolution with an infinite pair-entangled projected state (iPEPS). The alternating sequence is optimized with the final energy as a cost function. The energy is calculated with a tangent space power method for the sake of its stability.
The method is benchmarked in the 2D transverse field quantum Ising model near its quantum critical point against a ground state obtained by variational optimization of the iPEPS. The optimal BB sequence differs non-perturbatively from a sequence simulating quantum annealing or adiabatic preparation (AP) of the ground state. The optimal BB energy converges with the number of bangs much faster than the optimal AP energy.
\end{abstract}

\maketitle

\section{Introduction}
\label{sec:intro}

Understanding properties of strongly correlated quantum many-body systems is one of the long-standing problems in the theoretical/computational condensed matter physics, especially in two spatial dimensions where correlation effects are strong but, unlike in one-dimensional systems, integrability or numerically exact tractability are often missing. Exact diagonalization is limited to small system sizes by the exponential growth of the Hilbert space with the size of the system. Powerful Monte Carlo approaches are plagued by the notorious sign problem that can be circumvented by tensor networks for weakly entangled states. The entanglement is not a barrier for quantum simulators/computers but the present-day quantum hardware --- noisy intermediate scale quantum (NISQ) devices \cite{Preskill2018quantumcomputingin} --- can operate reliably for shallow circuits only.

In this work we employ a genuinely two-dimensional tensor product ansatz --- also known as the pair-entangled projected state (PEPS) \cite{nishino01, gendiar03, verstraete2004, Murg_finitePEPS_07,Cirac_iPEPS_08,Xiang_SU_08,Gu_TERG_08,Orus_CTM_09,
fu,Corboz_varopt_16, Vanderstraeten_varopt_16, Fishman_FPCTM_17, Xie_PEPScontr_17, Corboz_Eextrap_16, Corboz_FCLS_18, Rader_FCLS_18, Rams_xiD_18} --- to design the quantum approximate optimization algorithm (QAOA) \cite{farhi2014quantum}. The QAOA splits the target Hamiltonian into two non-commuting terms, $H=H_1+H_2$, and after initialization in a product state performs a sequence of unitary evolutions alternating between $H_1$ and $H_2$. This bang-bang (BB) \cite{Cerezo2021,PhysRevX.7.021027} sequence of time steps (or rotation angles) is optimized to minimize the final energy in the target Hamiltonian $H$ to obtain the best approximation to its ground state. It is preferable to minimize the number of BB steps that are equal to the depth of the quantum circuit. 
The shallowness of the allowed QAOA makes it ideally suited for classical simulation with tensor networks, as already demonstrated in 1D with matrix product states \cite{BB_MPS} (MPS). With the classical simulation, one can design an optimal BB protocol to prepare the ground state on a quantum hardware before it is subject to further quantum processing that goes beyond any classical simulation. In this work, we demonstrate that a similar tensor network method can be successfully employed on an infinite lattice in a two-dimensional system. 
This is not quite straightforward given the lack of tractable canonical structure, but see Ref. \onlinecite{Zaletel_iso}, which necessitates resorting to local updates in time evolution, like the neighborhood tensor update (NTU) \cite{ntu} used here, and to evaluate expectation values in controlled-approximation schemes. We calculate the final energy with a tangent space power method \cite{Vanderstraeten2019,tangent_Corboz} to warrant its stability for exotic BB sequences explored by the optimization algorithm, see App. \ref{ap:boundaries}. 

The rest of the paper is organized as follows. In Sec. \ref{sec:ntu} we outline the tensor network algorithm that we use to simulate the unitary evolution. The tangent space power method for expectation values is presented in some detail in App. \ref{ap:boundaries} and \ref{ap:expval}. In Sec. \ref{sec:sequences} we define the BB gate sequence as well as an alternative adiabatic preparation (AP) sequence that simulates quantum annealing \cite{annealing_review} with a digital quantum device. Numerical results for optimal sequences (both AP and BB) are presented in Sec. \ref{sec:results}. We conclude in Sec. \ref{sec:conclusion}.

\section{2D tensor network algorithm}
\label{sec:ntu}

\begin{figure}[t!]
\includegraphics[width=6cm]{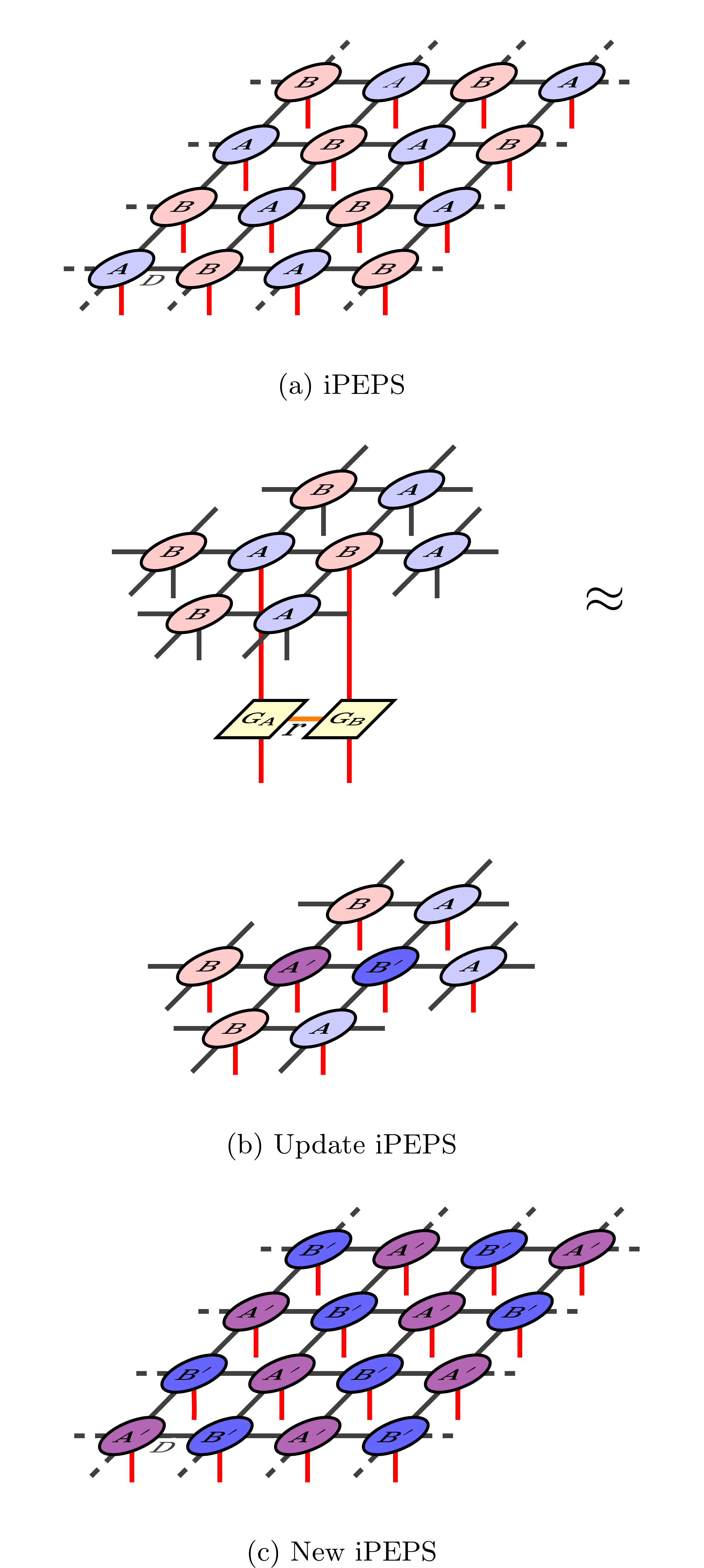} 
\figcaption{
{\bf NTU. }
In (a) infinite PEPS with tensors $A$ (purple) and $B$ (pink) on the two sub-lattices of an infinite checkerboard lattice. The red lines are physical spin indices and the black lines are bond indices, with bond dimension D, contracting NN site tensors. 
In one of Suzuki-Trotter steps a Trotter gate is applied to every NN pair of $A$-$B$ tensors along every horizontal row (but not to horizontal $B$-$A$ pairs). The gate can be represented by a contraction of two tensors, $G_A$ and $G_B$, by an index with dimension $r$. When the two tensors are absorbed into tensors $A$ and $B$ the bond dimension between them increases from $D$ to $rD$.
In (b) the $A$-$B$ pair -- with a Trotter gate applied to it -- is approximated by a pair of new tensors, $A'$ (deep purple) and $B'$ (darker blue), connected by an index with the original dimension $D$. The new tensors are optimized to minimize the difference between the two networks in (b).
After $A'$ and $B'$ are converged, they replace all tensors $A$ and $B$ in a new iPEPS shown in (c). 
Now the next Trotter gate can be applied.
}
\label{fig:NTU}
\end{figure}


Typical ground states of quantum many-body systems can be represented efficiently by tensor networks~\cite{Verstraete_review_08,Orus_review_14,Nishino_review_2022} including the matrix product states in one dimension (1D)~\cite{fannes1992}, the projected entangled pair state (PEPS) in 2D~\cite{Nishino_2DvarTN_04,verstraete2004} and 3D~\cite{Vlaar2021}, or the multi-scale entanglement renormalization ansatz (MERA)~\cite{Vidal_MERA_07,Vidal_MERA_08,Evenbly_branchMERA_14,Evenbly_branchMERAarea_14}. 
Recently an infinite PEPS ansatz (iPEPS) was employed to simulate unitary time evolution after a sudden Hamiltonian quench on infinite lattices\cite{CzarnikDziarmagaCorboz,HubigCirac,tJholeHubig,Abendschein08,SUlocalization,SUtimecrystal,ntu,mbl_ntu,BH2Dcorrelationspreading,ising2D_correlationsperading,schmitt2021quantum,Mazur_BH}. In this work, we use the neighbourhood tensor update (NTU) algorithm \cite{ntu} that was previously used to simulate the many-body localization \cite{mbl_ntu}, the Kibble-Zurek ramp in the Ising and Bose-Hubbard models \cite{schmitt2021quantum,Mazur_BH}, as well as thermal states obtained by imaginary time evolution in the fermionic Hubbard model~\cite{Hubbard_Sinha}.

Like in other schemes, in NTU the evolution operator is subject to the Suzuki-Trotter decomposition\cite{Trotter_59,Suzuki_66,Suzuki_76} into a product of one-site and nearest neighbor(NN) two-site Trotter gates. As each NN Trotter gate increases the bond dimension along its NN bond $r$ times, from $D$ to $rD$, the bond dimension has to be truncated back to a predefined $D$ to prevent its exponential growth with the number of gates. It has to be done in a way minimizing an error afflicted on the quantum state. There are several error measures, each of them implying a different algorithm: the simple update \cite{tJholeHubig,SUlocalization}, the full update \cite{fu,CzarnikDziarmagaCorboz}, the neighbourhood tensor update \cite{ntu,schmitt2021quantum,mbl_ntu}, or the gradient tensor update \cite{gradient}. The NTU error measure is explained in Fig. \ref{fig:NTU}. 
For each NN gate, the Frobenius norm of the difference between the left ($L$) and right ($R$) hand sides of Fig. \ref{fig:NTU} (b) is minimized. The NTU error $\delta_{i}$ of the $i^{\textrm{th}}$ gate is defined as the minimal norm $||L-R||$. $\delta_i$ is a rough estimate for an error inflicted on local observables by the bond dimension truncation. Accumulating truncation errors can eventually derail the time evolution. In the worst-case scenario, the errors are additive. This motivates a total NTU error \cite{Hubbard_Sinha}: 
\be 
\epsilon_{\textrm{NTU}} = \sum_i \delta_{i},
\label{eq:NTUerr}
\ee 
where the sum is over all performed NN Trotter gates.  

There are three main differences between this work and previous time evolution studies. 
The first is that, although the gate sequence has the same structure as in the Suzuki-Trotter decomposition, the gates are allowed arbitrary rotation angles instead of being restricted to small time steps. 
\changed{
The second is that evaluation of expectation values is done with an iPEPS boundary obtained by a tangent space power method \cite{Vanderstraeten2019}, see App. \ref{ap:boundaries}. It proved to be stable for exotic rotation sequences explored by the optimization algorithm. There is no reason to exclude sequences that seem too exotic as this is where we hope to find unexpected shortcuts to the target ground state. 
}
Finally, the third difference stems from a different motivation of the present study. Our aim is not just to find a tensor network state with minimal energy but also to design a gate sequence for a digital quantum computer. If we were just to find the state then we could accept significant NTU truncation errors as an inherent part of the algorithm targeting the minimal energy network, and in particular accept that the optimal gate sequence depends on the bond dimension. However, if we want the optimal gate sequence to reproduce the same state/energy when implemented on a digital quantum computer then we must suppress the gate truncation errors down to what can be safely considered as numerical zero.   

\section{Gate sequences}
\label{sec:sequences}

In this work, we consider the 2D transverse field quantum Ising Hamiltonian on an infinite square lattice: 
\be 
H = g H_1 + J H_2,
\ee 
where
\bea 
H_1 &=& -\sum_i X_i, \\
H_2 &=& -\sum_{\langle i,j \rangle} Z_i Z_j.
\eea 
Here each NN pair appears in the sum only once. $X_i$ and $Z_i$ stand for Pauli matrices $\sigma_x^i$ and $\sigma^z_i$, respectively. In the following we set the ferromagnetic coupling $J=1$. The model has a quantum phase transition at finite $g_c$ separating the ferromagnetic phase from the paramagnetic one. Its quantum Monte Carlo estimate is $g_c=3.04438(2)$~\cite{Deng_QIshc_02}. 

We want to prepare the ground state of the model at a given $g$ starting from an easy-to-prepare product state fully polarized along $X$ and then performing a finite number of steps, $N$. We employ two strategies. One is the real-time adiabatic quantum state preparation (AP) where the Hamiltonian is smoothly ramped from $H_1$ to final $H$ with the desired transverse field $g$. With $N$ steps allowed, the ramp is performed as a Suzuki-Trotter decomposition with a fixed time step, $\Delta t$. The step is a variational parameter, not necessarily small, optimized to minimize the energy at the end of the ramp. The other strategy is a bang-bang protocol (BB) where all rotation angles of the Suzuki-Trotter gates are free variational parameters. The bang-bang does not need to approximate the adiabatic quantum state preparation but it is allowed, and expected, to use the extra freedom to find its own short-cut towards the desired ground state. In this work, we target the ground state for $g=3.1$ which is close enough to the critical point to be challenging but still tractable by an iPEPS with limited bond dimension.

The real-time adiabatic quantum state preparation performs a smooth ramp of Hamiltonian parameters described by a function:
\be 
f(u)=\frac12\left[1+\sin(\pi(u-1/2))\right],
\ee 
parameterized by a time-like $u\in[0,1]$. The evolution operator is the second-order Suzuki-Trotter decomposition into $N$ time steps:
\bea 
U_{AP}\left( \Delta t \right) &=& e^{-\frac12 i\Delta t\cdot g H_1} \nonumber \\
&& e^{-        i\Delta t\cdot f\left[ (2N-1)/(2N) \right] H_2} \nonumber \\
&& e^{-        i\Delta t\cdot g H_1} \nonumber \\
&& \dots \nonumber \\
&& e^{-        i\Delta t\cdot g H_1} \nonumber \\
&& e^{-        i\Delta t\cdot f\left[1/(2N)\right] H_2} \nonumber \\
&& e^{-\frac12 i\Delta t\cdot g H_1}  
\label{eq:UA}
\eea 
with the time step $\Delta t$ being its only variational parameter that is not assumed small. As we initialize the system with an eigenstate of $H_1$ throughout this paper, the first step $\exp(-i\Delta t gH_1/2)$ does nothing but introduces a phase factor and thus can be dropped.
In contrast, the bang-bang evolution operator,
\bea 
&& U_{\textrm{BB}}\left(\beta_1,\dots,\alpha_N\right)=\nonumber \\
&& e^{-i\frac12 \alpha_N g H_1} \nonumber \\
&& e^{-i\beta_N H_2}            \nonumber \\
&& e^{-i\alpha_{N-1} g H_1}     \nonumber \\
&& \dots                        \nonumber \\
&& e^{-i\alpha_1 g H_1}         \nonumber \\
&& e^{-i\beta_1  H_2},          
\label{eq:UB}
\eea 
allows to optimize all $2N$ rotation angles $\beta_j,\alpha_j$ as free parameters. For comparison, the AP angles in \eqref{eq:UA} are constrained to be parameterized with a single time step $\Delta t$ as: 
\bea
\beta_j^{\textrm{(AP)}}    &=& \Delta t \cdot f\left[(2j-1)/(2N)\right], \nonumber \\
\alpha_{j}^{\textrm{(AP)}} &=& \Delta t.
\label{eq:alphabetaAP}
\eea 
The optimal BB angles may end up being close to their AP values, with perturbative differences meant to approximate the counter-adiabatic term\cite{del_Campo_2019} by the Suzuki-Trotter errors \cite{Wurtz2022counterdiabaticity}, or to follow a qualitatively different path along a non-perturbative short-cut to the desired ground state. 

\begin{figure}[t!]
\includegraphics[width=8.5cm]{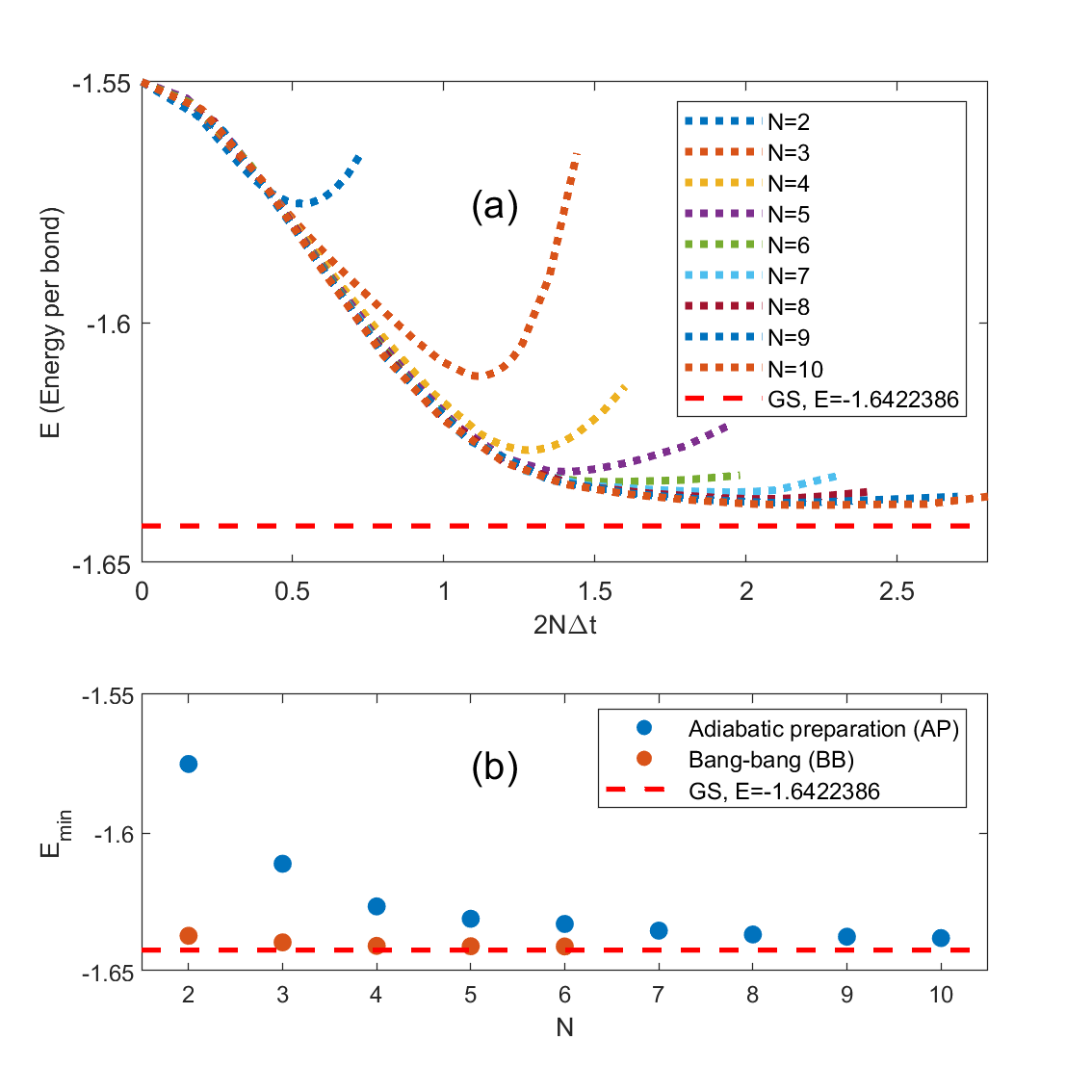}
\caption{
{\bf Adiabatic preparation (AP). ---}
In (a)
final energy at the end of the AP gate sequence in function of the total rotation angle $2N\Delta t$ for the transverse field $g=3.1$ and different quantum circuit depths $N$. The red dashed line is a benchmark ground state energy obtained by variational optimization \cite{corboz16b} of an iPEPS with $D=6$.
In (b)
the minimal final AP and BB energies in function of $N$.
}
\label{fig:AP_collapse}
\end{figure}

\begin{figure}[t!]
\includegraphics[width=8.5cm]{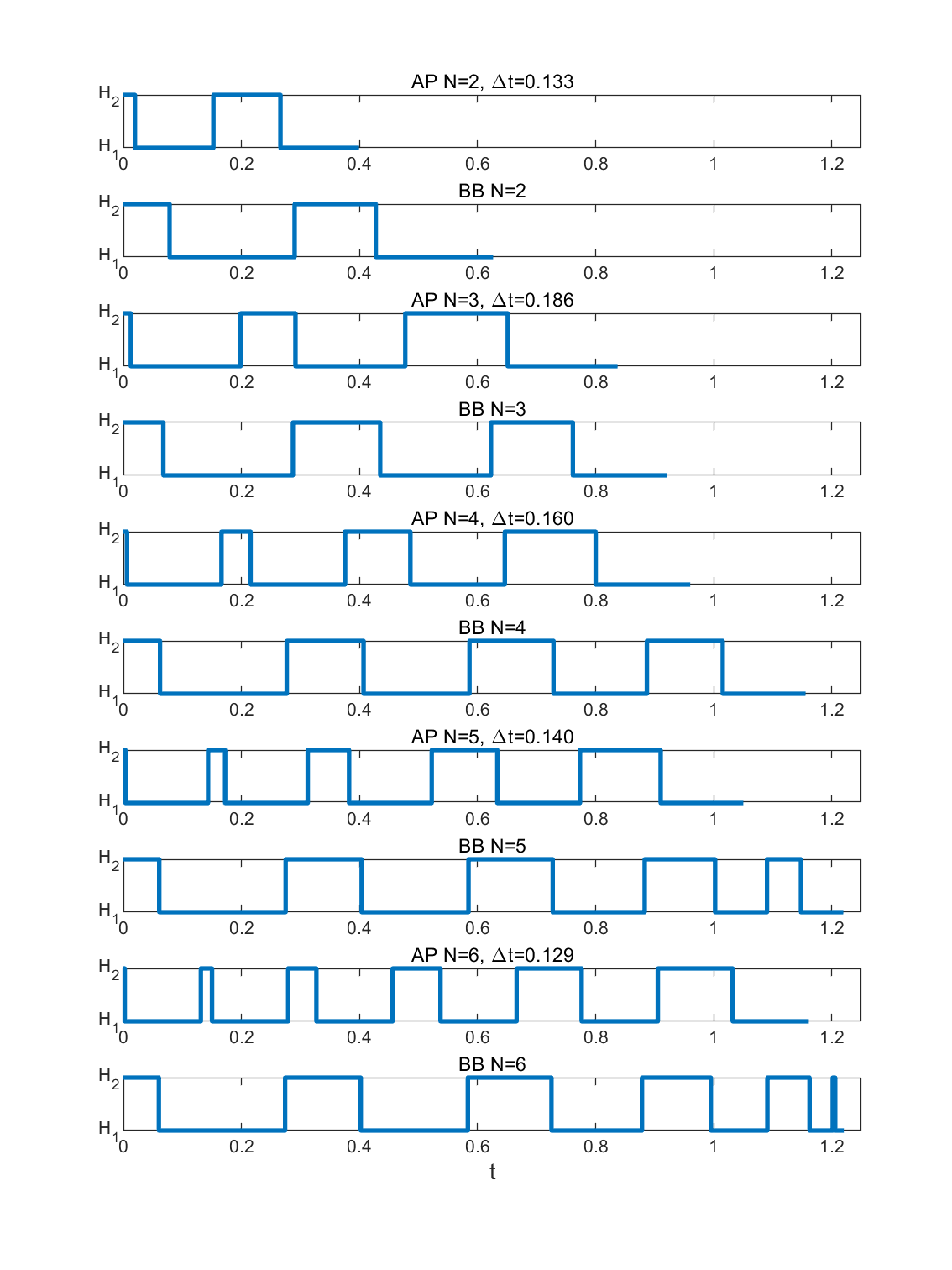}
\caption{{\bf Optimal gate sequences.}
The parameters $\beta_j,\alpha_j$ for the optimal bang-bang (BB) in \eqref{eq:UB} and the adiabatic quantum state preparation (AP) in \eqref{eq:UB} and \eqref{eq:alphabetaAP} for the optimal time step $\Delta t$. In both protocols, the unitary evolution is switching between the two-qubit Hamiltonian $H_2$ and the one-qubit $H_1$ with lengths of corresponding segments equal to $\beta_j$ and $\alpha_j$, respectively.
The optimal AP and BB sequences with the same number of steps $N$ differ non-perturbatively.   
}
\label{fig:patterns}
\end{figure}

\begin{figure}[t!]
\includegraphics[width=8.5cm]{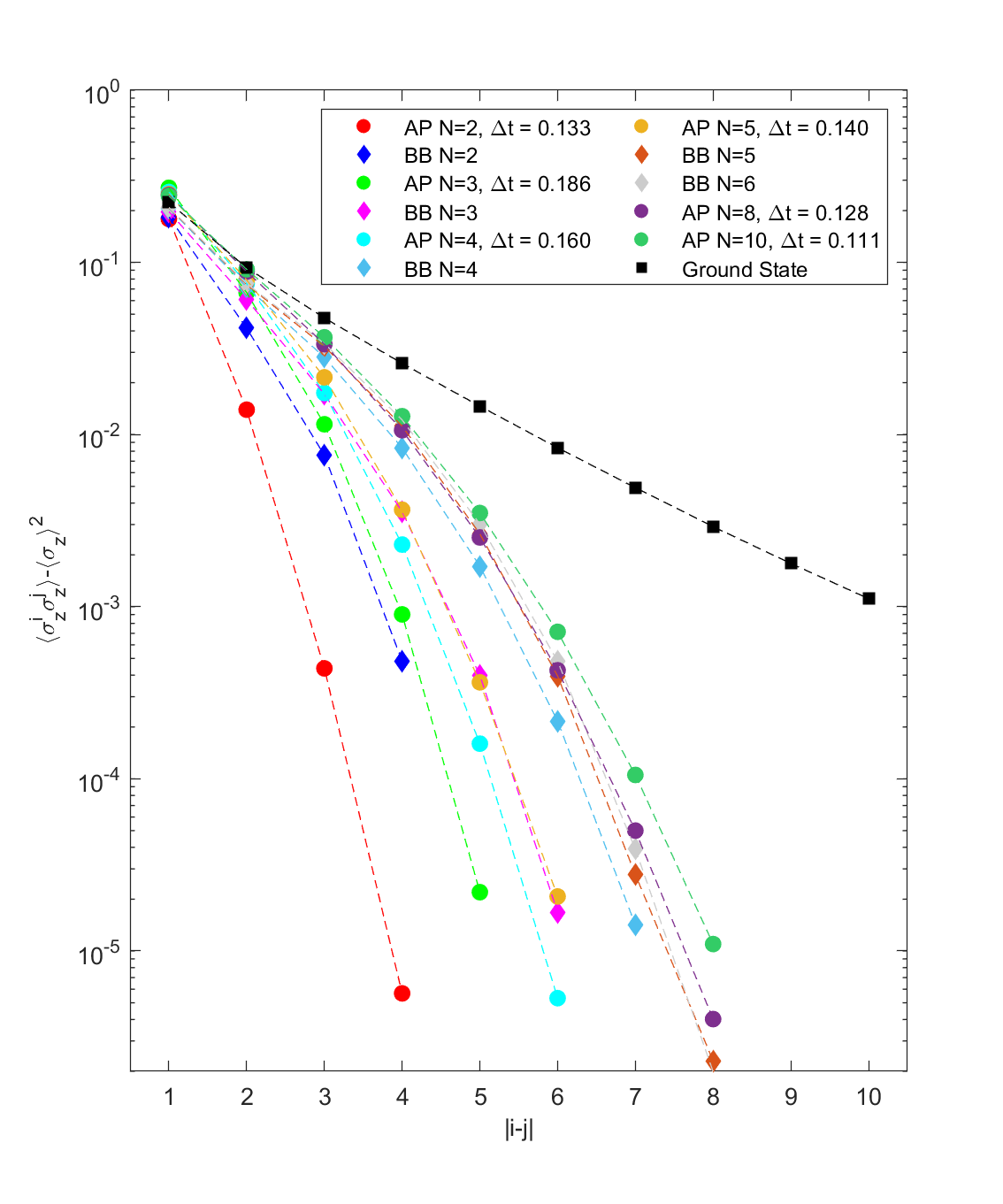}
\caption{{\bf Ferromagnetic correlator.} 
The connected part of the ferromagnetic correlation function is shown for the two types of protocols and for the ground state obtained variationally which serves as a benchmark. The ground state correlation length is $\xi=2.38$. Both the adiabatic preparation (AP) and the bang-bang (BB) improve towards the benchmark with increasing $N$. A finite circuit depth $N$ limits the maximal range of any non-zero correlations to $2N$. 
\label{fig:ZZ}
}
\end{figure}

\section{Results}
\label{sec:results}

Our aim is to achieve as good approximation to the ground state of $H$ as possible in a limited number of steps $N$ that defines the depth of the quantum circuit. For the relatively shallow circuits that we consider it was enough to use iPEPS bond dimension $D=8$. The largest total NTU errors \eqref{eq:NTUerr} that we encountered for different $N$ are listed in Table \ref{tab:results}. They are small enough for the obtained BB patterns to be transferred to a quantum computer without any modification to mitigate the NTU errors.
Calculation of expectation values required another bond dimension, $\chi$, that is a refinement parameter for approximate contraction of an infinite squared norm of the iPEPS state, see appendix \ref{ap:expval}. In this study, $\chi=40$ proved to be sufficient. 

\begin{table}[h!]
\centering
\begin{tabularx}{0.45\textwidth} { 
  | >{\centering\arraybackslash}X| 
  | >{\centering\arraybackslash}X| 
  | >{\centering\arraybackslash}X|
  | >{\centering\arraybackslash}X|}
 \hline
 N & $\epsilon_{\textrm{NTU}}$ & $E_\textrm{AP}$ & $E_\textrm{BB}$ \\
 \hline
 \hline
 2  & $0$                 & $-1.575331$ & $-1.637082$ \\
 \hline
 3  & $0$                 & $-1.611126$ & $-1.639453$\\
 \hline
 4  & $5.1\times 10^{-7}$ & $-1.626484$ & $-1.64071$\\
 \hline
 5  & $1.6\times 10^{-6}$ & $-1.630946$ &  $-1.64085$ \\
 \hline
 6  & $1.9\times 10^{-6}$ & $-1.633067$ & $-1.64091$ \\
 \hline
 7  & $1.2\times 10^{-8}$ & $-1.635215$ &  $-$\\
 \hline
 8  & $2.0\times 10^{-8}$ & $-1.636623$ &  $-$\\
 \hline
 9  & $3.2\times 10^{-8}$ & $-1.637411$ &  $-$\\
 \hline
 10 & $5.6\times 10^{-7}$ & $-1.637896$ & $-$ \\
 \hline
 \hline
 variational & $-$ & $-1.6422386$ & $-1.6422386$  \\
\hline
\end{tabularx}
\caption{
{\bf Summary of results. }
$N$ is the depth of the quantum circuit, i.e., the number of layers of NN gates applied to the initial product state. The second column lists corresponding maximal total NTU errors \eqref{eq:NTUerr} encountered for the optimal AP or BB gate sequences. As the SVD-rank of the NN gate is $r=2$, for $D=2^3$ there is no error up to $N=3$. The third and the fourth column list the optimal energies per bond for the transverse field $g=3.1$. The ground state energy obtained by variational optimization\cite{corboz16b} with $D=6$ is shown as a benchmark.   
} 
\label{tab:results}
\end{table}

We consider the adiabatic preparation first. When $\Delta t$ is small enough then the Suzuki-Trotter (ST) errors become negligible and all that matters is only the overall ramp time $N\cdot\Delta t$. Figure \ref{fig:AP_collapse} (a) shows the energy at the end of the ramp in function of the total rotation angle $2N\cdot\Delta t$ for several values of $N$. For small $\Delta t$ the plots with different $N$ collapse demonstrating the accuracy of the ST approximation in this regime. This collapse fails when $\Delta t$ is too long for the ST decomposition because $g\cdot \Delta t\ll1$ or $J\cdot \Delta t\ll1$. For the considered $g=3.1$ and $J=1$ the former condition is stronger and fails first. Increasing $\Delta t$ further makes the ST errors larger and, consequently, makes the final energy grow with $\Delta t$. For each $N$ there is optimal $\Delta t\propto 1/g$ when the final energy is minimal. 

The energy gap at $g=3.1$ between the ground state and the first excited state is $\Delta=0.15$, see appendix in \onlinecite{schmitt2021quantum}, which means that the minimal time required to make the straightforward AP to be adiabatic is $N\cdot\Delta t\approx 2\pi/\Delta\approx40$. In this situation, the final energy begins to decay exponentially with the total time. With $g\cdot \Delta t\ll1$ necessary to tame the ST errors, the minimal $N$ corresponding to the minimal time is $N\approx120$. This is the condition for the final AP energy to begin to decay exponentially with the total annealing time $N\cdot\Delta t$. 

In Figure \ref{fig:AP_collapse} (b) we compare the optimal AP energies with the optimal BB ones. In the BB protocol the gate angles $\alpha$ and $\beta$ were optimized with two built-in algorithms of {\verb|MATLAB|}: one is the local minimum searching algorithm {\verb|fmincon|}; the other is the global minimum searching algorithm \verb|patternsearch|\cite{patternsearch,MATLAB}. Instead of using the global optimizer directly, we first feed {\verb|fmincon|} initially with the bang-bang pattern of the best AP approach and then use the \verb|fmincon|. It returns a local minimum which can serve as a benchmark for the global minimum. Then, we use this local minimum as the initial input to \verb|patternsearch|. In this way, \verb|patternsearch| can return a global minimum that is smaller than the local one. Thanks to the extra freedom in the choice of rotation angles, the convergence of the optimized energy with $N$ is much faster for the BB than the AP sequence. The optimal AP and BB gate sequences for $N=2..6$ are shown in Fig. \ref{fig:patterns}. We can see that for each $N$ the BB sequence is non-perturbatively different from the AP one. The optimal BB sequence for $N=2..4$ comes out the same no matter if the optimization is initialized with the optimal AP sequence or a random one. 
For $N=5,6$ we were forced to change the strategy and initialize the BB sequence with the optimal BB sequence for $N-1$. This explains the similarity between the BB patterns for $N=4$ and $N=5$: $N=5$ pattern differs from $N=4$ by an extra final $H_2$ gate. The BB patterns for $N=5,6$ are even closer: $N=6$ has an extra spike of $H_2$. Nevertheless, the energy continues to decrease with $N$.

In both AP and BB we use the final energy as a convenient cost function. The local observable is relatively easy to compute but it cannot stand for full characterization of the ground state near the quantum critical point with long ferromagnetic correlations. Fig. \ref{fig:ZZ} shows ferromagnetic correlators after several AP and BB sequences together with the one obtained by a variational optimization\cite{corboz16b} that serves here as a benchmark. All AP and BB correlators have a shorter range than the variational one. This is understandable as they have higher energy but also because the range of any non-zero correlator is limited to at most $2N$ by construction. A BB correlator is closer to the benchmark than its AP counterpart with the same $N$ as the BB state has lower energy. At $N=6$ BB becomes comparable to AP with $N=10$ at a medium range though its far tail remains lower, as may be explained by the limitation imposed by the finite $N$. The BB procedure is doing a better job than the AP in the sense of achieving lower energy and stronger short-range correlations for a smaller $N$ even though the same small $N$ limits its correlation range. 

As a final remark, in App. \ref{ap:p2f} we attempt to prepare a ground state on the ferromagnetic side of the quantum phase transition with the same minimal set of gates, $X$ and $ZZ$, and starting from the same $X$-polarized initial state.  

\section{Conclusion}
\label{sec:conclusion}

We provided a proof of principle that the iPEPS time evolution combined with tangent space methods can be used to design optimal shallow bang-bang quantum gate sequences preparing the ground state of a 2D quantum lattice Hamiltonian. The BB sequences converge with the number of gates faster than sequences simulating the adiabatic quantum state preparation thanks to a far bigger number of variational parameters allowing for a non-perturbative short-cut towards the target state. 

\changed{
The infinite system considered here is a convenient benchmark, as it requires only two sublattice tensors, but finite systems can also be treated by PEPS with different tensors on different lattice sites~\onlinecite{METTS_ising_Hubbard,king2024computational}.
Moreover, the translationally invariant sequence optimized on an infinite lattice is also a good starting point for a finite system as it is already optimal in its bulk, i.e. farther away from the lattice's edges than the correlation length in the final state. The rest can be readily improved by optimizing only the gates on the edge. As the iPEPS is correlated, this optimization would improve the final state not only on the edge itself but also up to the correlation length distance from it. 
}

The figure data can be downloaded from \url{https://uj.rodbuk.pl/dataset.xhtml?persistentId=doi%3A10.57903%2FUJ%2FBSBOMY}. 
\acknowledgements
%
%
This research was supported in part by the National Science Centre (NCN), Poland under project 2019/35/B/ST3/01028 (J.D.) and
project 2021/03/Y/ST2/00184 within the QuantERA II Programme that has received funding from the European Union Horizon 2020 research and innovation programme under Grant Agreement No 101017733 (Y.Z.).
The research was also supported by a grant from the Priority Research Area DigiWorld under the Strategic Programme Excellence Initiative at Jagiellonian University.

\bibliography{KZref.bib} 
\appendix

\tikzset{TA/.style  = {shape=rectangle, ,draw=blue!50, fill=blue!20, minimum size=8mm}}
\tikzset{TB/.style  = {shape=rectangle, ,draw=red!50, fill=red!20, minimum size=8mm}}
\tikzset{A/.style  = {shape=circle, ,draw=blue!50, fill=blue!20, minimum size=8mm}}
\tikzset{B/.style  = {shape=circle, ,draw=red!50, fill=red!20, minimum size=8mm}}
\tikzset{C/.style  = {shape=rectangle, ,draw=green!50, fill=green!20, minimum size=8mm}}
\tikzset{O/.style  = {shape=rectangle, ,draw=black!100, fill=yellow!20, minimum size=8mm}}
\begin{figure}[t!]
    \centering
    \includegraphics[width=8cm]{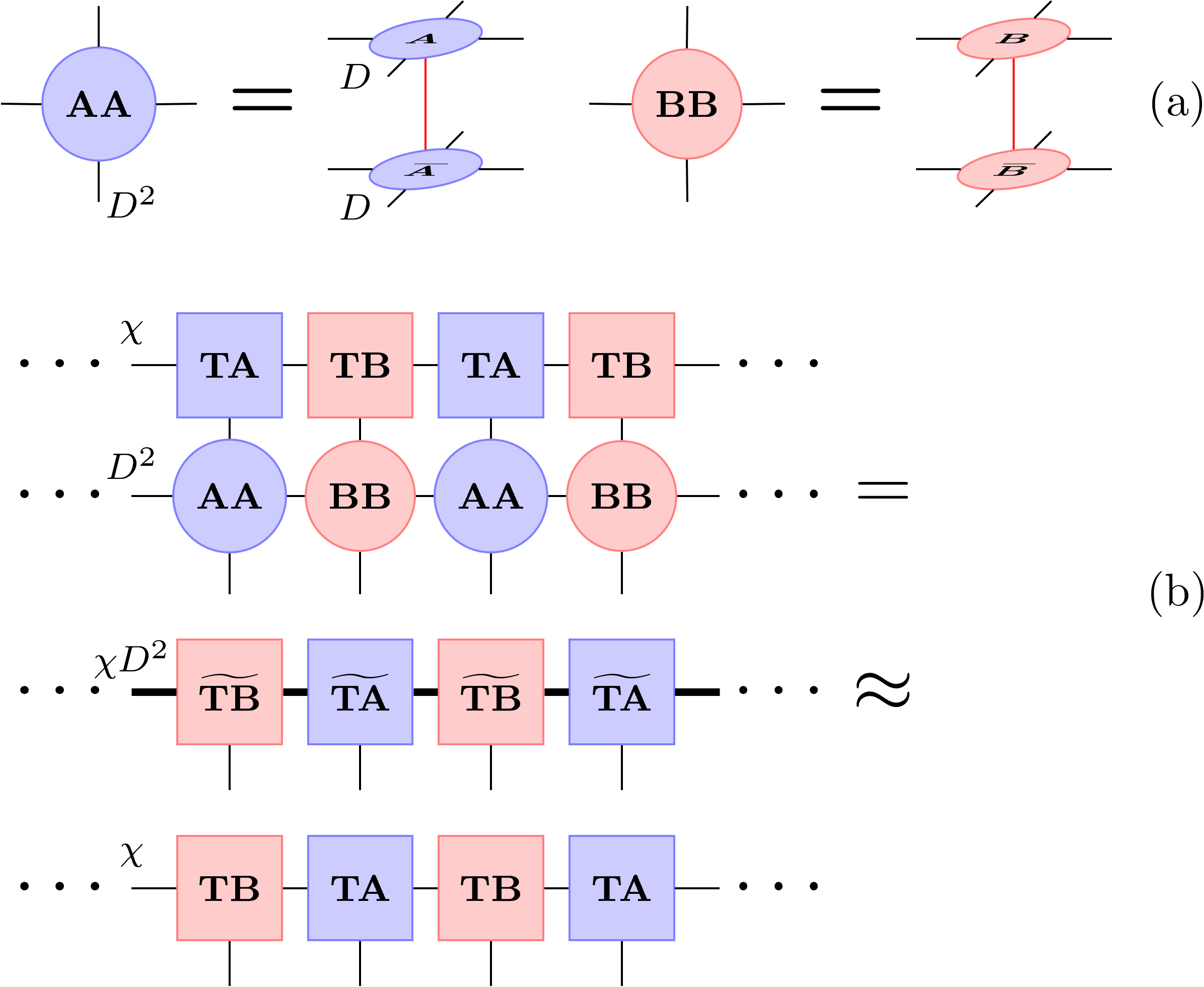}
    \caption{\textbf{Boundary MPS ansatz. ---} 
    In (a)
    double PEPS tensor $\textrm{AA}$ ($\textrm{BB}$) consists of iPEPS tensor $A$ ($B$) contracted through the physical index with its complex conjugate $\overline{A}$ ($\overline{B}$).
    In (b)
    the top boundary MPS consists of tensors $\textrm{TA}$ and $\textrm{TB}$.    
    A row of tensors $\textrm{AA}$ and $\textrm{BB}$ makes a row transfer matrix. 
    After the boundary is applied with the row transfer matrix it becomes a new boundary MPS 
    made of new tensors $\widetilde{\textrm{TA}}$ and $\widetilde{\textrm{TB}}$ with bond dimension $\chi D^2>\chi$.
    The dimension is compressed back by approximating the new MPS with an MPS made of new tensors $\textrm{TA}$ and $\textrm{TB}$ with bond dimension $\chi$.  }
    \label{fig:boundarymps}
\end{figure}

\begin{figure}[t!]
\centering
\includegraphics[width=8cm]{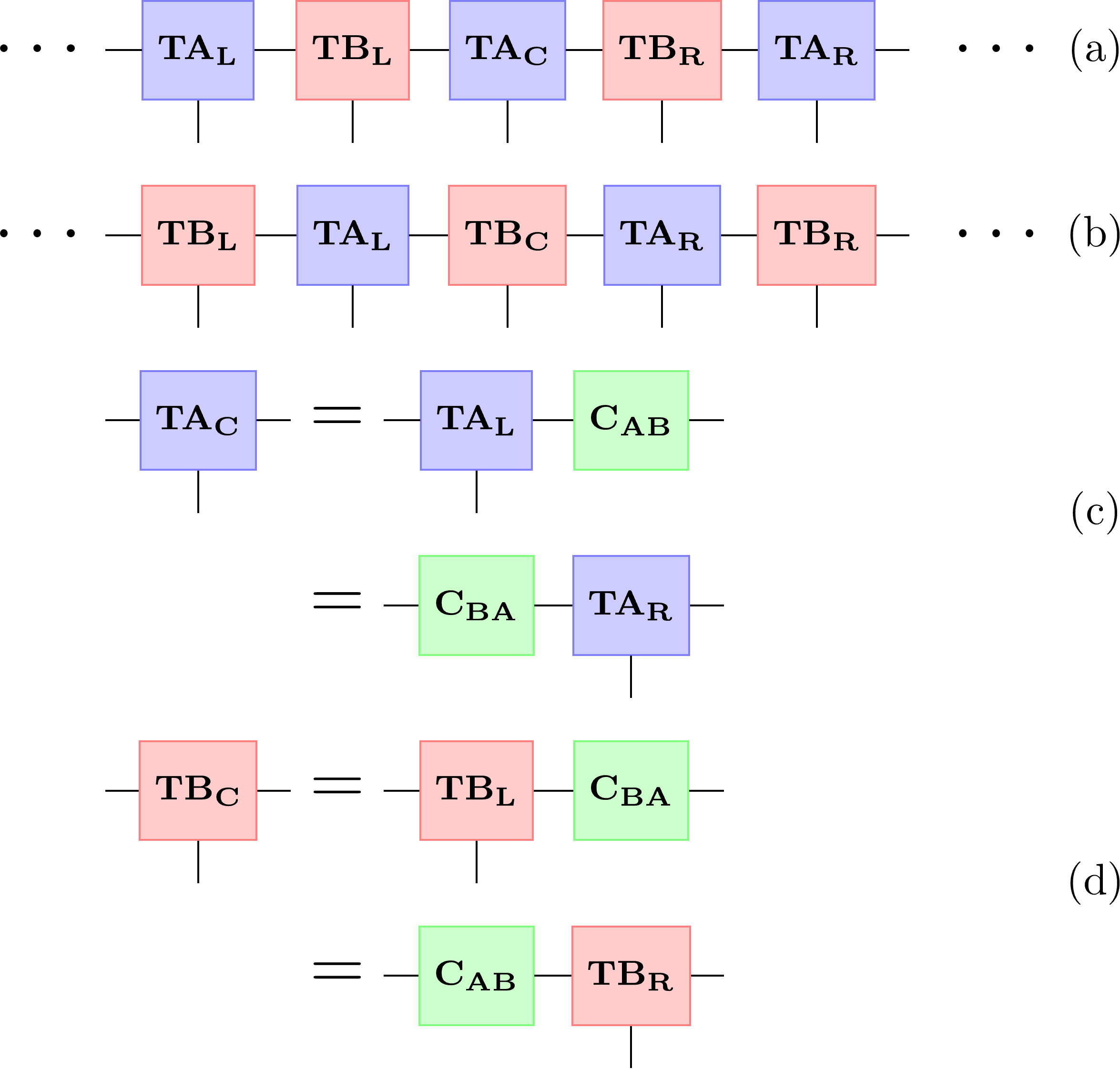}
                    
\caption{
\textbf{Canonical forms. ---} 
In (a,b) the mixed canonical boundary MPS with $\textrm{TA}_C$/$\textrm{TB}_C$ as the canonical centers.
Here $\textrm{TA}_L$/$\textrm{TA}_R$ and $\textrm{TB}_L$/$\textrm{TB}_R$ are left/right canonical tensors. 
The forms (a,b) are equivalent thanks to the relations in (c,d).
The canonical centers and $C_{AB}/C_{BA}$ are normalized to $1$. }
\label{fig:CanonicalForm}
\end{figure} 

\begin{figure}[t!]
\includegraphics[width=8cm]{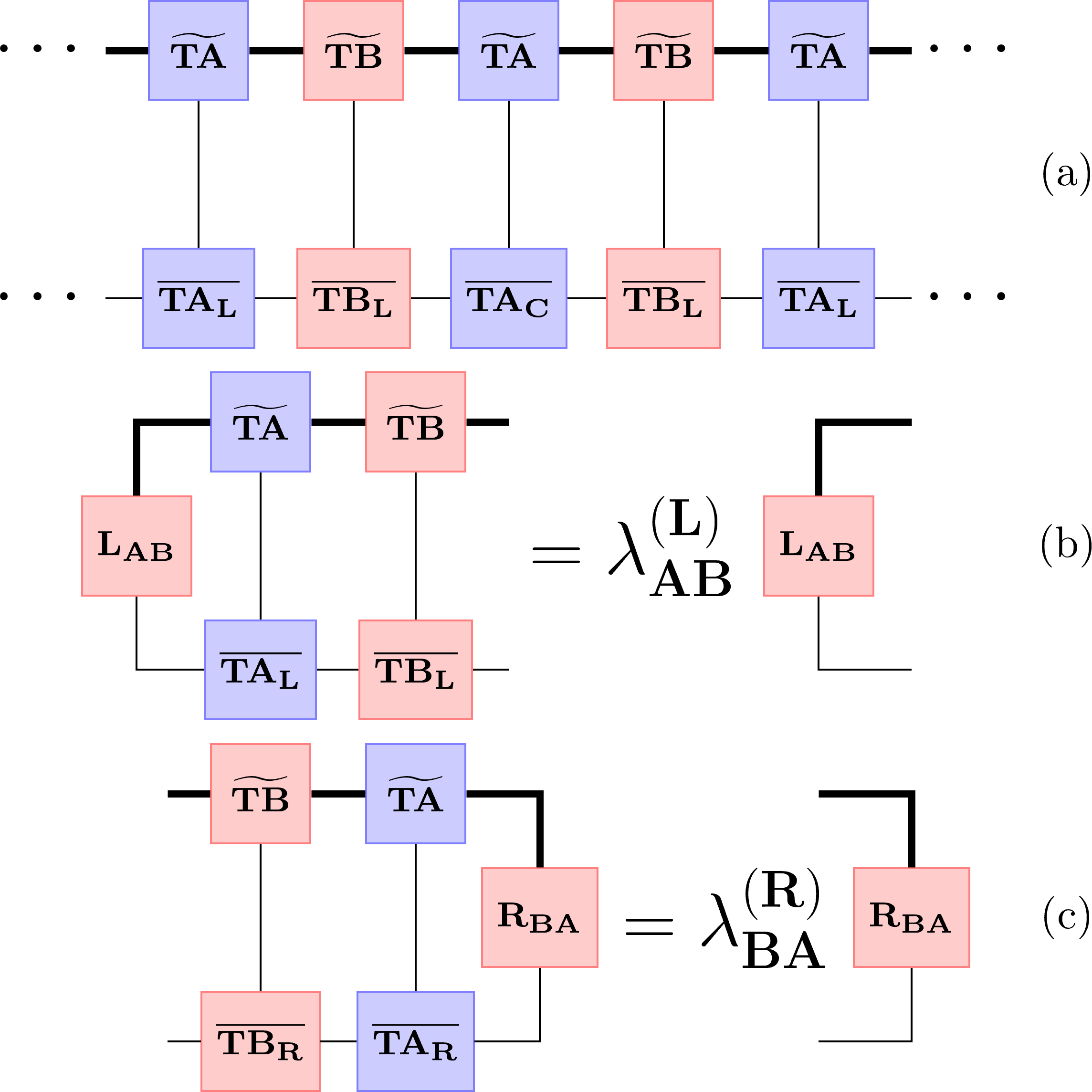}
\caption{
\textbf{Overlap.}
The overlap between the new $\chi D^2$-iMPS boundary and the compressed $\chi$-iMPS boundary in Fig. \ref{fig:boundarymps} (b) is shown here in (a). Its left part is a semi-infinite product of two transfer matrices, $\widetilde{TA}-\overline{TA_L}$ and $\widetilde{TB}-\overline{TB_L}$, where the overline means a complex conjugation. 
Therefore, the left part can be replaced by the left leading eigenvector of their product defined in (b). 
Similarly, the right part can be replaced by the leading right eigenvector in (c). }
\label{fig:Overlap}
\end{figure}

\begin{figure}[t!]
    \centering
    \includegraphics[width=4cm]{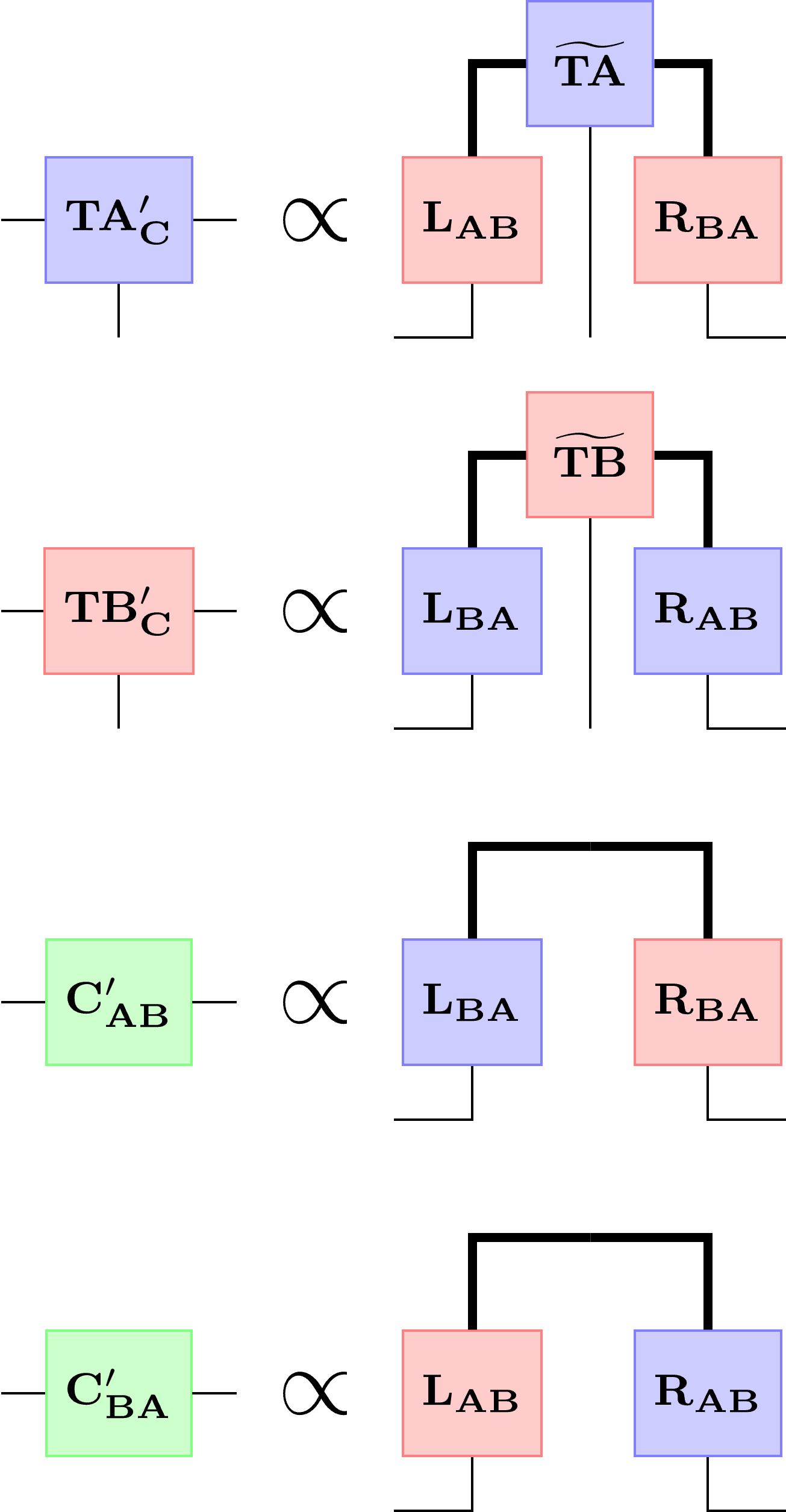}
    \caption{
    \textbf{Canonical centers' update.} 
    New canonical centers $\textrm{TA}_C'$ and $\textrm{TB}_C'$ as well as $C_{AB}'$ and $C_{BA}'$ are updated with the left and right leading eigenvectors defined in Fig. \ref{fig:Overlap} (b,c). 
    \label{fig:UpdateCanonicalCenterl}}
\end{figure}

\begin{figure}[t!]
    \centering
    \includegraphics[width=6.0440cm]{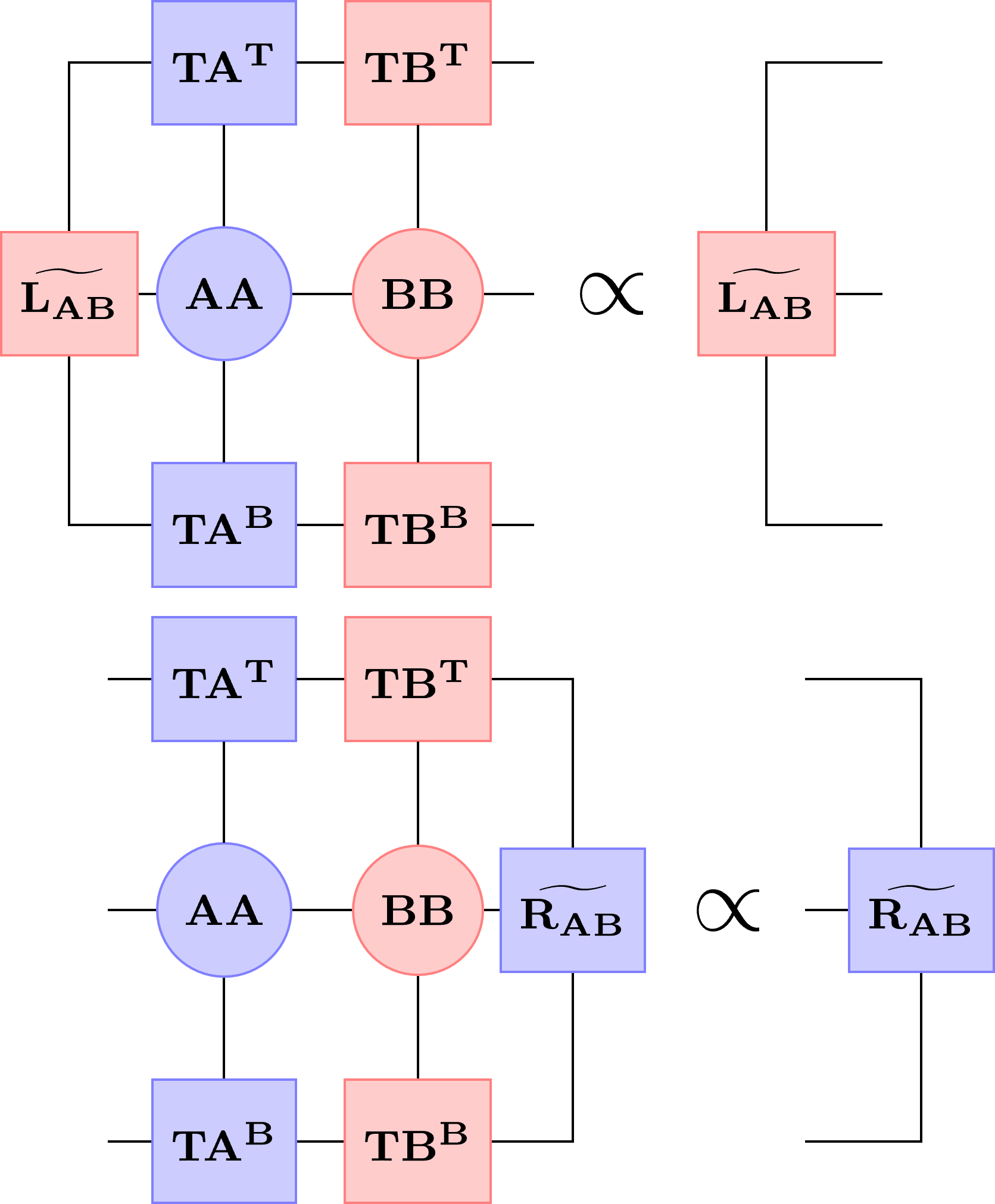}
    \caption{
    \textbf{Fixed points.}
    The top and bottom iPEPS boundaries with a row transfer matrix in between are an infinite product of transfer matrices. Here we introduce their leading left and right eigenvectors: $\widetilde L_{AB}$ and $\widetilde R_{AB}$ to be used in Fig. \ref{fig:Expecationvalue} below.
    }
    \label{fig:fixed2}
\end{figure}

\section{iPEPS boundaries}
\label{ap:boundaries}

We calculate observables with a variational tangent space MPS technique \cite{Vanderstraeten2019}. Here we summarize the algorithm that we used for the considered checkerboard lattice. 

In Fig. \ref{fig:boundarymps} (a) we define the double iPEPS tensors that occur in the squared norm of the iPEPS, \braket{\psi}{\psi}, where the bra and the ket are, respectively, the conjugate iPEPS and the iPEPS itself. Approximate row-by-row contraction of the squared norm from top to bottom results in an upper boundary represented by an infinite matrix product state (iMPS) with ``physical'' indices with dimension $D^2$ and bond indices with dimension $\chi$, see Fig. \ref{fig:boundarymps} (b). $\chi$ is a refinement parameter controlling the accuracy of approximations made during the row-by-row contraction. The contraction is done by repeated application of a row transfer matrix to the boundary iMPS until its convergence, see Fig. \ref{fig:boundarymps} (b). After every application, the bond dimension of the boundary iMPS increases from $\chi$ to $\chi D^2$ and has to be compressed back to the original $\chi$ to avoid its divergence. The compression is done by approximating the iMPS with a new iMPS with bond dimension $\chi$. The $\chi$-iMPS is optimized by maximizing its overlap with the $\chi D^2$-iMPS with the help of the tangent space methods \cite{Vanderstraeten2019,tangent_Corboz}.

They represent the boundary iMPS by its mixed-canonical forms in Fig. \ref{fig:CanonicalForm} (a,b). The tensors to the left/right of the canonical center are left/right isometries. With QR decomposition in Fig. \ref{fig:CanonicalForm} (c,d) the mixed-canonical iMPS can be brought to an equivalent form with a central bond matrix $C_{AB}$ or $C_{BA}$. Furthermore, the equations in panels (c,d) allow one to move the orthogonality center along the chain without changing the state represented by the iMPS. They warrant that the mixed canonical MPS has a hidden translational symmetry.

The overlap between the new $\chi D^2$-iMPS boundary and the new compressed $\chi$-iMPS boundary is shown in Fig. \ref{fig:Overlap} (a). The orthogonality center at site $A$ divides the diagram into two semi-infinite parts. The left one can be interpreted as a semi-infinite alternating product of two transfer matrices: $\widetilde{TA}-\overline{TA_L}$ and $\widetilde{TB}-\overline{TB_L}$. The right one is a product of the other two transfer matrices: $\widetilde{TA}-\overline{TA_R}$ and $\widetilde{TB}-\overline{TB_R}$. Therefore, the left/right part can be replaced by leading left/right eigenvectors of the left/right transfer matrices defined in \ref{fig:Overlap} (b,c). After convergence of the maximized overlap the leading eigenvalues should become the same, $\lambda^{(L)}_{AB}=\lambda^{(L)}_{BA}=\lambda^{(R)}_{AB}=\lambda^{(R)}_{BA}$, and their common magnitude should achieve its maximal value.

Using the left and right leading eigenvectors, new canonical centers $\textrm{TA}_C'$ and $\textrm{TB}_C'$ as well as new $C_{AB}'$ and $C_{BA}'$ are calculated as shown in Fig. \ref{fig:UpdateCanonicalCenterl}. New left/right isometries, $TA'_{L,R}$ and $TB'_{L,R}$, are to be updated in such a way that they satisfy the relations in Fig. \ref{fig:CanonicalForm} (c,d) as closely as possible and, therefore, we require them to minimize four cost functions:
\bea 
&&
\left|\left| TA'_C - C'_{BA} \cdot TA'_R \right|\right|^2, \label{eq:1} \\
&&
\left|\left| TA'_C - TA'_L \cdot C'_{AB} \right|\right|^2, \\
&&
|| TB'_C - TB'_L \cdot C'_{BA} ||^2,~~\\
&&
|| TB'_C - C'_{AB} \cdot TB'_R ||^2.
\eea 
Here all tensors were reshaped into matrix forms and $\cdot$ means a matrix multiplication. Following \citet{Vanderstraeten2019}, we make polar decompositions:
\bea 
TA'_C &=& U^{L}_{AC} \cdot P^L_{AC} = P^R_{AC} \cdot U^R_{AC}, \\
TB'_C &=& U^{L}_{BC} \cdot P^L_{BC} = P^R_{BC} \cdot U^R_{BC}, \\
C'_{AB} &=& U^L_{AB} \cdot P^L_{AB} = P^R_{AB} \cdot U^R_{AB}, \\
C'_{BA} &=& U^L_{BA} \cdot P^L_{BA} = P^R_{BA} \cdot U^R_{BA}. 
\eea 
The isometries, $U^{L/R}_{AC/BC}$ and $U^{L/R}_{AB/BA}$, are used to update the left and right isometric tensors as
\bea 
TA'_L &=& U^L_{AC} \cdot U^{L\dag}_{AB}, ~~TB'_L = U^L_{BC} \cdot U^{L\dag}_{BA}, \\
TA'_R &=& U^{R\dag}_{BA} \cdot U^R_{AC}, ~~TB'_R = U^{R\dag}_{AB} \cdot U^R_{BC}. \label{eq:8}
\eea 
The whole procedure in Fig. \ref{fig:UpdateCanonicalCenterl} followed by Eqs. from \eqref{eq:1} to \eqref{eq:8} is repeated until convergence. 

After the convergence of the overlap the next row transfer matrix is applied and again an overlap between the resulting $\chi D^2$-iMPS and a new compressed $\chi$-iMPS is maximized iteratively. The row transfer matrices are applied repeatedly until convergence of the upper iMPS boundary. The lower boundary and, if necessary, left and right boundaries are obtained similarly.  

\begin{figure}[t!]
    \centering
\includegraphics[width=8.3445cm]{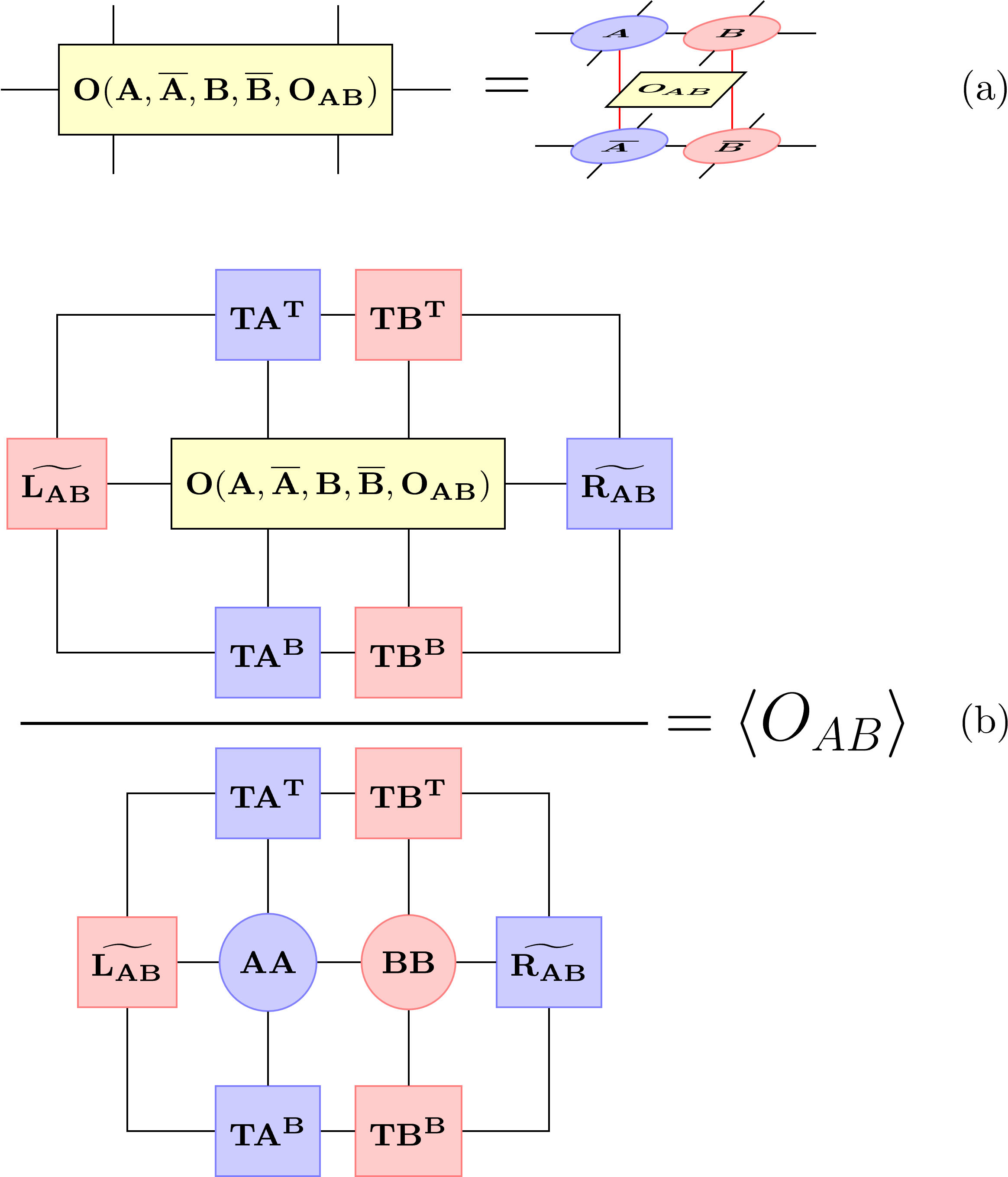}
    \caption{
    \textbf{Expectation value calculation. ---} 
    In (a) a two-site operator $O_{AB}$ is placed between the iPEPS and its conjugate.
    In (b) the left and right fixed points from Fig. \ref{fig:fixed2} together with the top and bottom iPEPS boundary tensors are combined to obtain the expectation value of $O_{AB}$.  }
    \label{fig:Expecationvalue}
\end{figure}

\section{Expectation values}
\label{ap:expval}

With the top and bottom iPEPS boundaries, we can calculate expectation values as shown in Figs. \ref{fig:fixed2} and \ref{fig:Expecationvalue}. To begin with, we place the row transfer matrix between the two boundaries and obtain an infinite product of the transfer matrices. The infinite product can be replaced by their leading left and right eigenvectors shown in Fig. \ref{fig:fixed2}. These fixed points are employed in Fig. \ref{fig:Expecationvalue} to obtain an expectation value of an operator $O_{AB}$ supported on two nearest-neighbor sites. Expectation values for operators with support along the vertical direction can be calculated in a similar fashion.

\changed{
\section{BB to the ferromagnetic phase}
\label{ap:p2f}

In this appendix we start from the same initial product state fully polarized along $X$ but this time we target the ground state on the ferromagnetic side of the quantum phase transition. In order to isolate problems arising from crossing the critical point from those due to representing correlations in the state, here we consider zero final transverse field with the fully polarized product ferromagnetic ground state.

AP employs a ramp of Hamiltonian parameters described by a function:
\be 
f(u)=\frac12\left(1+u|u|\right),
\label{eq:f(u)_p2f}
\ee 
parameterized by a time-like $u\in[-1,1]$. The Hamiltonian 
\be  
H(u) = f(u) H_2 + \tilde f(u) g_c H_1,
\ee 
where $\tilde{f}=1-f$, interpolates between the transverse-field $H_1$ and the ferromagnetic $H_2$. Its time-dependence slows near the critical point at $u=0$ to reduce quantum Kibble-Zurek excitations~\cite{QKZ1,QKZ2,QKZ3,d2005,d2010-a,d2010-b,Z-d}. At the beginning and the end it has discontinuous time derivatives that add some excitations but they are negligible when compared to the leading KZ excitations due to crossing the critical point. Given that the function is bound to be discretized with a limited number of time steps $N$, it may not be flexible enough to attempt nullifying the initial and final derivatives together with slowing at the critical point.

The $N$-step Suzuki-Trotter decomposition of the AP evolution operator is
\bea 
U_{AP}\left( \Delta t \right) &=&
   e^{-\frac12 i\Delta t\cdot \tilde{f}\left[ 1 \right] g_c H_1} \nonumber \\
&& e^{-        i\Delta t\cdot f\left[ (2N-1)/(2N) \right] H_2} \nonumber \\
&& e^{-        i\Delta t\cdot \tilde{f}\left[ (2N-2)/(2N) \right] g_cH_1} \nonumber \\
&& \dots \nonumber \\
&& e^{-        i\Delta t\cdot \tilde{f}\left[ 2/(2N) \right] g_c H_1} \nonumber \\
&& e^{-        i\Delta t\cdot f\left[1/(2N)\right] H_2} \nonumber \\
&& e^{-\frac12 i\Delta t\cdot \tilde{f}\left[ 0 \right] g_c H_1}. 
\label{eq:UA_p2f}
\eea 
with the time step $\Delta t$ being its only variational parameter. The action of the first gate on the initial state is trivial and can be ignored. The BB evolution operator is 
\bea 
&& U_{\textrm{BB}}\left(\beta_1,\dots,\alpha_N\right)=\nonumber \\
&& e^{-i\frac12 \alpha_N g_c H_1} \nonumber \\
&& e^{-i\beta_N H_2}            \nonumber \\
&& e^{-i\alpha_{N-1} g_c H_1}     \nonumber \\
&& \dots                        \nonumber \\
&& e^{-i\alpha_1 g_c H_1}         \nonumber \\
&& e^{-i\beta_1  H_2}.          
\label{eq:UB_p2f}
\eea 
All its $2N$ rotation angles $\beta_j,\alpha_j$ as free parameters. 
Table \ref{tab:results_p2f} contains the optimal AP and BB energies together with the corresponding NTU errors. 

It turns out that, with the same minimal set of $X$ and $ZZ$ gates, it is much harder to cross to the ferromagnetic phase than to prepare a paramagnetic ground state. These two gates respect the $Z_2$ symmetry that is spontaneously broken at the phase transition. They cannot generate a symmetry breaking bias, like e. g. the simple one in Ref. \onlinecite{QKZteor-r,KZexp-x}, to make the passage through the critical point more adiabatic. Nevertheless, the extra freedom of the BB ansatz gives it a clear advantage over AP.
Including an extra $Z$-gate, that breaks the $Z_2$ symmetry, could generate a symmetry breaking bias, the same or more general than in Ref. \onlinecite{QKZteor-r}, to open a spectral gap when passing across the critical point.
Alternatively, including an extra $Y$-gate, that also breaks the $Z_2$ symmetry, an initial rotation around $Y$ could transform the initial $X$-polarized state into a $Z$-polarized one and then --- via the Kramers-Wannier duality --- a dual version of the optimal sequence in Fig. \ref{fig:patterns} would prepare a ferromagnetic ground state dual to the paramagnetic one prepared in the main text.  
\begin{figure}[t!]
\includegraphics[width=8.5cm]{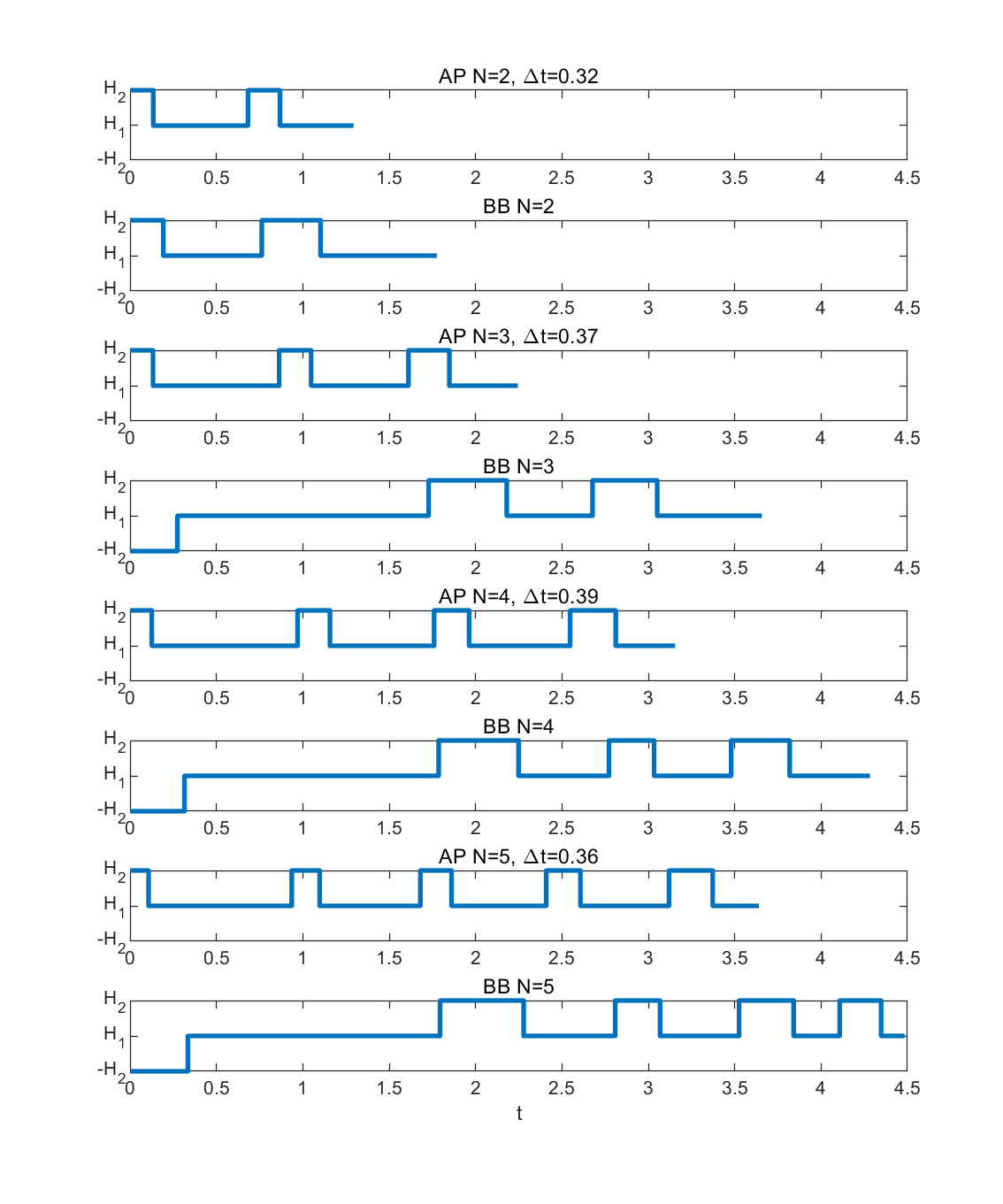}
\caption{\textbf{Optimal gate sequences of BB to the ferromagnetic phase.} --- Sets of parameters $\beta_j$ and $\alpha_j$ for the optimal bang-bang (BB) in (\ref{eq:UB}) that tune the system from the paramagnetic phase to the ferromagnetic phase. The energy of optimized BB2 is already lower than the energy of the adiabatic approach (AP) with $N=5$ (see Table \ref{tab:results_p2f}). Notice that for BB3, BB4 and BB5, we have one bang that corresponds to a gate of the form $\exp(-i\beta_j(-H_2))$ in which $\beta_j>0$. This ``time-reversing'' gate is significantly different from the gates in the adiabatic preparation approaches (see (\ref{eq:f(u)_p2f}) and (\ref{eq:UA_p2f})), manifesting the non-adiabatic nature of the bang-bang approach. }
\label{fig:patterns}
\end{figure}
}
\begin{table}[b]
\centering
\begin{tabularx}{0.45\textwidth} { 
  | >{\centering\arraybackslash}X| 
  | >{\centering\arraybackslash}X| 
  | >{\centering\arraybackslash}X|
  | >{\centering\arraybackslash}X|}
 \hline
 N & $\epsilon_{\textrm{NTU}}$ & $E_\textrm{AP}$ & $E_\textrm{BB}$ \\
 \hline
 \hline
 2  & $0$  & $-0.28202$ & $-0.48065$ \\
 \hline
 3  & $0$  & $-0.32152$ & $-0.52906$\\
 \hline
 4  & $2.1\times 10^{-4}$ & $-0.35180$ & $-0.6015(8)$\\
 \hline
 5  & $5.1\times 10^{-4}$ & $-0.42315$ & $-0.616(48)$ \\
 \hline
 \hline
 exact & $-$ & $-1$ & $-1$  \\
\hline
\end{tabularx}
\caption{
{\bf Summary of results: para to ferro. }
$N$ is the depth of the quantum circuit. The second column lists corresponding maximal total NTU errors \eqref{eq:NTUerr} encountered for the optimal AP or BB gate sequences. The third and the fourth columns list the optimal energies per bond for the transverse field $g=0$. The exact ground state energy is $-1$. 
} 
\label{tab:results_p2f}
\end{table}

\end{document}